-------------------------------------------------------------------------------
\documentstyle[12pt]{article}
\begin{document}
\baselineskip = 24pt

\begin{titlepage}
\vspace{2.5cm}
\begin{center}{\large \vspace{4.0cm} \bf The quantum bialgebra associated with
the eight-vertex R-matrix}\\
\vspace{1cm}
D.B.Uglov \footnotemark  \\
\footnotetext{e-mail: denis@max.physics.sunysb.edu}
\vspace{1cm}
Department of Physics, State University of New York at Stony Brook \\
Stony Brook, NY 11794-3800, USA \\
\vspace{0.5cm}
February 24, 1993
\vspace{3cm}

\begin{abstract}
The quantum bialgebra related to the Baxter's eight-vertex R-matrix is found as
a quantum deformation of the Lie algebra of $sl(2)$-valued automorphic
functions on a complex torus.
\end{abstract}

\end{center}
\end{titlepage}

{\bf 1.} Recent years witnessed an extensive developement of the theory of
quantum groups and their quasiclassical limits - Lie bialgebras . Although
since the work of Belavin and Drinfel'd [3] it is known, that solutions of the
Classical Yang-Baxter equation can be classified into three categories: the
rational, the trigonometric and the elliptic ones, the main developement took
place in the rational and the trigonometric cases. The underlying algebraic
structures in these cases were found to be the affine Kac-Moody Lie algebras
(classical case)and the Yangians and the quantum affine Kac-Moody algebras
(quantum case) [5,6]. One of the major developements in the elliptic case was
the discovery of the Sklyanin algebra [4]. This algebra, however, has no
coproduct and thus it is not a bialgebra. Another important work had been done
by Reyman and Semenov-Tyan-Shanskii [2], who found the Lie bialgebras
associated with the elliptic solutions of the Classical Yang-Baxter equation.
The simplest
    example of their alge
bras is the Lie algebra of $sl(2)$-valued automorphic meromorphic functions on
a complex torus. The Lie bialgebra structure of this algebra is given by the
classical r-matrix of the Landau-Lifshitz model [2]. In the work [7] the
generators and the defining relations of this Lie algebra had been found. In
the present letter we address the problem of quantization of the above Lie
bialgebra. The result is a quantum bialgebra related to the eight-vertex
R-matrix. We use the term `` quantum bialgebra`` to designate a Hopf algebra
without the antipode.

{\bf 2.} In [7] it is shown, that the Lie algebra of $sl(2)$-valued automorphic
meromorphic functions on a complex torus which we denote by ${\cal
E}_{k,\nu^{\pm}}$ is a finitely generated (infinite-dimensional) {\bf C}-Lie
algebra defined upon the six generators $ \{ x^{\pm}_{k} \}_{k=1,2,3} $ by the
relations ( $ i,j,k $ below is any cyclic permutation of 1,2,3 ):

\begin{equation}
[x_{i}^{\pm},[x_{j}^{\pm},x_{k}^{\pm}]] = 0 ,
\end{equation}
\begin{equation}
[x_{i}^{\pm},[x_{i}^{\pm},x_{k}^{\pm}]]-[x_{j}^{\pm},[x_{j}^{\pm},x_{k}^{\pm}]]= J_{ij}x_{k}^{\pm},
\end{equation}
\begin{equation}
[x_{i}^+,x_{i}^{-}]=0 ,
\end{equation}
\begin{equation}
[x_{i}^{\pm},x_{j}^{\mp}] = \sqrt{-1}( w_{i}(\nu^{\mp}-\nu^{\pm})
x_{k}^{\mp}-w_{j}(\nu^{\mp}-\nu^{\pm}) x_{k}^{\pm}.
\end{equation}

, where  $ \nu^{+},\nu^{-} \in {\rm {\bf T}} = {\rm {\bf C}}/({\rm {\bf Z}} 4K
+ {\rm {\bf Z}} 4iK^{'}), \;  \nu^{+}-\nu^{-} \neq {\rm {\bf Z}} 2K + {\rm {\bf
Z}}2iK^{'}  ; K , K^{'} $-are the complete elliptic integrals of the moduli $k$
and $k^{'}$ correspondingly: $k^{2}+k^{'2}=1 ;\; J_{12}=k^{2},\:
J_{23}=k^{'2},\: J_{31}=-1;$ and $ w_{1}(u)=\frac{1}{sn(u,k)} ;\;
w_{2}(u)=\frac{dn}{sn}(u,k) ;\; w_{3}(u)=\frac{cn}{sn}(u,k) , u\in {\rm {\bf
T}}.$

The structure of a Lie bialgebra upon ${\cal E}_{k,\nu^{\pm}}$ is defined by
the classical r-matrix of the Landau-Lifshitz model [2]. A certain
trigonometric limit $k \rightarrow 0 $ of ${\cal E}_{k,\nu^{\pm}}$ coincides
with the loop algebra $A_{1}^{(1)'}/(centre)$ [7].

{\bf 3.} To define a quantum deformation of the Lie algebra (1-4) we introduce
a deformation parameter $\eta \in {\rm {\bf C}}$ and recall the definition of
the Baxter's eight-vertex R-matrix [1]:  $ R(u)=I \otimes I + \sum_{k=1}^{3}
\frac{w_{k}(u+i\eta)}{w_{k}(i\eta)} \sigma_{k} \otimes \sigma_{k} \; \in
Mat_{2}\otimes Mat_{2},\; u \in {\rm {\bf T}} $ , $ I $ is $ 2 \times 2 $
identity matrix , and $ \sigma_{k} $ are the Pauli matrices. Let $H =
R(0)R^{'}(0)$.

Introduce associative {\bf C}-algebra ${\cal E}_{k,\nu^{\pm},\eta}$ generated
by the elements:  {\bf1} (identity) ,$ T_{ab}^{\pm},\overline{T}_{ab}^{\pm}, \;
a,b \in \{ 1,2 \} $. The defining relations of $ {\cal E}_{k,\nu^{\pm},\eta} $
have the following form:
\begin{equation}
\overline{T}_{1}^{\pm}(H_{12}-H_{14})T_{1}^{\pm} +
\overline{T}_{3}^{\pm}(H_{34}-H_{32})T_{3}^{\pm} +
T_{2}^{\pm}(H_{23}-H_{21})\overline{T}_{2}^{\pm} +
T_{4}^{\pm}(H_{41}-H_{43})\overline{T}_{4}^{\pm} = 0 ,
\end{equation}
\begin{equation}
\overline{T}_{1}^{\pm}T_{1}^{\pm}=T_{1}^{\pm}\overline{T}_{1}^{\pm}={\rm
{\bf1}}I_{1},
\end{equation}
\begin{equation}
R_{12}(\nu^{\mp}-\nu^{\pm})T_{1}^{\pm}T_{2}^{\mp}=T_{2}^{\mp}T_{1}^{\pm}R_{12}(\nu^{\mp}-\nu^{\pm}) .
\end{equation}
We adopt the standard convention: $ T_{n}^{\pm} $ means a matrix with $ {\cal
E}_{k,\nu^{\pm},\eta} $ -valued entries , which acts nontrivially only in the
$n-$th factor of $ {\rm {\bf C}}^{\otimes^{m}} $ ( $ m = 4 $ in (5) , $ m = 2 $
in (7) ) and coincides there with $ T_{ab}^{\pm} $.

{\bf PROPOSITION} \begin{it}
${\cal E}_{k,\nu^{\pm},\eta}$ is a bialgebra, i.e. there exist a coproduct $
\Delta : {\cal E}_{k,\nu^{\pm},\eta} \rightarrow {\cal E}_{k,\nu^{\pm},\eta}
\otimes {\cal E}_{k,\nu^{\pm},\eta} $ and a counit $ \varepsilon : {\cal
E}_{k,\nu^{\pm},\eta} \rightarrow {\bf {\rm C }}$, such, that:
\begin{equation}
\Delta(ab) = \Delta(a) \Delta(b) , \; \;  \; a , b \in {\cal
E}_{k,\nu^{\pm},\eta}\end{equation}
\begin{equation}
(\varepsilon \otimes id)\Delta(a) =  (id \otimes \varepsilon)\Delta(a) = a , \;
\; a \in {\cal E}_{k,\nu^{\pm},\eta}.
\end{equation} \end{it}

Explicit expressions for $\Delta $, and $ \varepsilon $ are given by the
following formulae:
\begin{eqnarray}
\Delta (T_{ab}^{\pm}) = \sum_{c=1}^{3} T_{ac}^{\pm} \otimes T_{cb}^{\pm}, \; \;
\Delta (\overline{T}_{ab}^{\pm}) = \sum_{c=1}^{3} \overline{T}_{cb}^{\pm}
\otimes \overline{T}_{ac}^{\pm},\; \; \Delta ({\rm {\bf1}})={\rm {\bf1}}
\otimes {\rm \bf{1}}.
\end{eqnarray}
\begin{eqnarray}
\varepsilon (T_{ab}^{\pm}) = \delta_{ab}, \; \varepsilon
(\overline{T}_{ab}^{\pm}) = \delta_{ab}, \; \; \varepsilon({\rm {\bf 1}})= 1 .
\end{eqnarray}

In the quasiclassical limit $ \eta \rightarrow 0 $ , which is described as
follows:
\begin{eqnarray}
T^{\pm} = I + 2i\eta \sum_{k=1}^{3} x_{k}^{\pm}\sigma_{k} + O(\eta^{2}),\;
\overline{T}^{\pm} = I + 2i\eta \sum_{k=1}^{3} \overline{x}_{k}^{\pm}\sigma_{k}
+ O(\eta^{2}),
\end{eqnarray}
one recovers from the defining relations (5-7) the defining relations (1-4) for
the generators $ \{x_{k}^{\pm} \}_{k=1,2,3} $ in $ {\cal E}_{k,\nu^{\pm}} $ .
Note, that from (6) it follows that $ \overline{x}_{k}^{\pm} = -x_{k}^{\pm} $.

The eight-vertex R-matrix $ R(u-v) $ appears again as an intertwiner between
the tensor products $ \pi_{u} \otimes \pi_{v} $ and $ \pi_{v} \otimes \pi_{u} $
of the simplest 2-dimensional representations $ \pi_{u(v)} $ of $ {\cal
E}_{k,\nu^{\pm},\eta} $ parametrized by $ u,v \in {\rm {\bf T}} $ :
\begin{eqnarray}
\pi_{u}(T^{\pm}) = \frac{1}{\sqrt{D(u-\nu^{\pm})}}R(u-\nu^{\pm}),\;
\pi_{u}(\overline{T}^{\pm}) = \frac{1}{\sqrt{D(u-\nu^{\pm})}}R(-u+\nu^{\pm}),
\end{eqnarray}
\begin{equation}
D(u) = 1-\sum_{k=1}^{3} \frac{w_{k}(u+i\eta)w_{k}(u-i\eta)}{w_{k}^{2}(i\eta)}.
\end{equation}

To establish a connection between the elliptic and the trigonometric cases we
need to perform the trigonometric limit $k \rightarrow 0$ of ${\cal
E}_{k,\nu^{\pm},\eta}$. By analogy with the classical case [7], we do this
limit at $\nu^{+}=i\frac{3}{2}K^{'}, \; \nu^{-}=i\frac{1}{2}K^{'}$. The author
had succeeded to recover from (5-7) under this limit all the defining relations
of $U_{q}(A_{1}^{(1)'}/(centre))$ at $q=e^{2\eta}$ except the quantum Serre
relations.

{\bf Acknowlegement} \\
The author is grateful to I.T.Ivanov and L.A.Takhtadjan for  helpful
discussions. \\
\vspace{1cm}

\begin{large}
{\bf References} \end{large} \\
1. Baxter, R. J., {\em Ann. Phys. } {\bf 76}, 1 (1973).\\
2. Reyman, A. G. and Semenov-Tyan-Shanskii, M. A., {\em Journ. Sov. Math.} {\bf
46}, 1631 (1989). \\
3. Belavin, A. A. and Drinfel'd V. G., {\em Funct. Anal. Appl. } {\bf 17}, 220
(1984).\\
4. Sklyanin, E. K., {\em Funct. Anal. Appl.} {\bf 16}, 263 (1983).\\
5. Drinfel'd V. G., {\em Proceedings of the ICM}, Berkeley ,CA U.S.A., 798
(1986). \\
6. Chari, V. and Pressley, A., {\em L'Enseignement Math\'{e}matique} {\bf 36},
267 (1990); {Comm. Math. Phys.} {\bf 142}, 261 (1991).\\
7. Uglov, D. B., to be published.
\end{document}